\documentclass[dvips]{article}
\usepackage{icrctc07}
\def\mincir{\raise -2.truept\hbox{\rlap{\hbox{$\sim$}}\raise5.truept \hbox{$<$}\ }}
\def\mincireq{\hbox{\raise0.5ex\hbox{$<\lower1.06ex\hbox{$\kern-1.07em{\sim}$}$}}}
\def\magcir{\raise-2.truept\hbox{\rlap{\hbox{$\sim$}}\raise5.truept \hbox{$>$}\ }}

\title{Gamma-ray blazars: the combined AGILE and MAGIC views}
\shorttitle{$\gamma$-ray blazars: AGILE and MAGIC}

\authors{M.~Persic$^{1,3}$, A.~De~Angelis$^{2}$, F.~Longo$^{3}$, M.~Tavani$^{4}$}
\shortauthors{Persic et~al.}
\afiliations{
$^1$INAF-Trieste, Trieste, Italy \\
$^2$University of Udine and INFN, Udine, Italy \\ 
$^3$INFN-Trieste, Trieste, Italy \\
$^4$INAF-Roma, Rome, Italy }
\email{persic@oats.inaf.it}

\abstract{ We describe the emission properties of blazars, i.e. the AGNs 
that, due to their peculiar orientation w.r.t. the observer, allow the 
most penetrating and direct view of their central engine. After showing 
that the extragalactic GeV-TeV sky is dominated by blazars of various types, 
we discuss the kind of blazars that are likely to be jointly detected by 
AGILE and MAGIC. }

\begin{document}
\maketitle

\section{Blazars}

Supermassive black holes (SMBHs) reside in the cores of most galaxies. 
The fueling of SMBHs by infalling matter produces the spectacular activity 
observed in active galactic nuclei (AGNs).

The current AGN paradigm includes a central engine, most likely a SMBH, surrounded by an
accretion disk and by fast-moving clouds, which emit Doppler-broadened lines (\cite{up95,p07}). 
More distant clouds emit narrower lines. Absorbing material in some flattened configuration 
(e.g., a torus) obscures the central parts, so that for transverse lines of sight only the 
narrow-line emitting clouds are seen and the source is classified as a so-called 'Type 2' 
AGN. The near-IR to soft-X-ray nuclear continuum and broad lines, including the UV bump 
typical of classical quasars, are seen only when the source is viewed face-on: the object 
is then classified a 'Type 1' AGN. In radio-loud objects, which constitute $\sim$10$\%$ of 
all AGNs, the infalling matter switches on powerful collimated jets which shoot out from the 
SMBH in opposite directions, likely perpendicular to the disk, at relativistic speeds. The 
origin of such jets is one of the fundamental open issues in astrophysics.

If a relativistic jet is viewed at small angle to its axis the observed emission is 
amplified by relativistic beaming (Doppler boosting and aberration)
	\footnote{ Defining the relativistic Doppler factor as 
	$\delta$$\equiv$$[\Gamma$$(1$$-$$\beta \cos \theta)]^{-1}$ (with 
	$\beta$$=$$v/c$ the jet speed normalized to the speed of light and 
	$\Gamma$$=$$1/$$\sqrt{(1-\beta^2)}$, and $\theta$ the angle w.r.t. the
	line of sight), the observed and intrinsic luminosities are related 
	by $L_\nu^{\rm obs}$$=$$\delta^p$$L_\nu^{\rm em}$ 
	with $p$$\sim$2-3, and the variability timescales are related 
	by $\Delta t_{\rm obs}$$=$$\delta^{-1}$$\Delta 
	t_{\rm em}$. For $\theta$$\sim$0$^{\circ}$ and $\delta$$\sim$2$\,\Gamma$ 
	the observed luminosity can be amplified by factors 400 -- 10,000 
	(for, typically, $\Gamma$$\sim$10 and $p$$\sim$2-3); whereas 
	$\theta$$\sim$$1/\Gamma$ implies $\delta$$\sim$$\Gamma$, 
	with a luminosity amplification of $\sim$100--1,000.},
allowing a deep insight into the physical conditions and emission processes prevailing in 
relativistic jets. Sources in which the jet (due to its favorable orientation) dominates 
the observed emission are called blazars. Although a minority among AGNs, blazars are of 
paramount importance for studying the physics of relativistic jets. Blazars are characterized 
by a non-thermal continuum spanning the radio to $\gamma$-ray band. 
The multiwavelength emission of blazars is strongly and rapidly variable, mostly at high 
energies, and it carries information on the timescales of particle acceleration and energy 
losses and on the dynamics of the emitting region.

The spectral energy distributions (SEDs) of blazars are generally characterized by two broad 
humps, peaking at, respectively, IR/X-ray and GeV-TeV frequencies (\cite{u+97}). Analyses of 
blazar SEDs (\cite{f+98,g+98}) have shown that: 
{\it (i)} higher-luminosity objects have both humps peaking at lower frequencies; 
{\it (ii)} the luminosity ratio between the high-energy component and the low-energy component 
increases with luminosity; 
{\it (iii)} at the highest luminosities the $\gamma$-ray output dominates the total luminosity. 

The SEDs of blazars are mainly interpreted as arising from the co-spatial synchrotron emission 
(peaked in the IR/X-ray range), from a time-varying population of ultra-relativistic electrons 
moving in a strong magnetic field, and the Compton emission of the synchrotron photons upscattered 
by the parent electrons into the $\gamma$-ray regime (synchrotron-self-Compton [SSC] model (\cite{j+74}). 
The emitting electrons are accelerated within the relativistic jets which transport energy from 
the central SMBH to the external regions of extragalactic radiosources (\cite{r67}). 
This highly complex physics is approximated, for the purpose of modelling the observed SEDs, 
with a series of relativistically moving homogeneous regions (blobs), where particle acceleration 
and radiation take place (\cite{br78,bk79,m80,k81,k89,r82,g+85,gm89,m+92}). The high-energy 
emission, with its extremely fast and correlated multifrequency variability, indicates 
that often a single region dominates the emission.
Within the SSC framework, the spectral humps are mostly interpreted as the synchrotron peak and Compton 
peak (e.g., \cite{p+98, k+02}). 
%
%

HE (multi-MeV to multi-GeV) and VHE (multi-GeV to multi-TeV) data are of crucial importance 
to close the SSC model. Even in the simplest one-zone synchro-self-Compton model of blazar 
emission, knowledge of the whole SED up to the VHE regime is 
required for a complete description of the emitting electrons' distribution and environment 
(e.g., (\cite{t+98,bp99,k+99}). Specifically, the parameters that specify the properties of 
the emitting plasma in the basic SSC model are: the electron distributions normalization, low- 
and high-energy slopes, and min/break/max energy, and the plasma blobs magnetic field, size 
and Lorentz factor. Knowing only the IR/X-ray peak would give info on the shape (i.e., the 
broken-power-law slopes) of the electron distribution but would leave all other parameters 
unconstrained (e.g., \cite{t+98}). 

Blazar observations are therefore a top priority for VHE astrophysics. Simultaneous broadband 
$\gamma$-ray data (obtained in the HE range with the space-borne AGILE and GLAST telescopes, 
and in the VHE range with the ground-based IACTs), in conjunction with X-ray data (from, e.g., 
Suzaku, Chandra, XMM, Swift) will allow us to more accurately know the parameters describing 
the population of emitting electrons.

\section{The GeV and TeV skies}

Before discussing the perspective HE-VHE AGILE-MAGIC observations we need to assess the present 
status of the $\gamma$-ray sky. 

At HE frequencies, the 3rd EGRET catalogue \cite{har99} includes 271 sources, $\magcir$130 of which 
are now known to be extragalactic (\cite{p07}; many sources are still unidentified), all of which are 
AGNs (except the Large Magellanic Cloud), $\sim$97$\%$ of which are blazars. Of these, $\sim$93$\%$ 
are low-frequency--peaked blazars (LBLs) and only $\sim$3$\%$ are high-frequency--peaked blazars 
(HBLs). Therefore, blazars -- mostly of the LBL type -- dominate EGRET's extragalactic HE sky. 

Blazars dominate (currently, 16/17 TeV sources) the extragalactic VHE sky, too. However, these are mostly HBLs 
-- at present, only 1/16 is of the LBL type. The HBL dominance descends from a selection bias: for a 
given HE flux, HBLs have a higher VHE flux than LBLs, because both spectral humps are shifted to 
higher frequencies.

After concluding that the broadband (HE+VHE) $\gamma$-ray sky is dominated by blazars, one comment 
in order. Given their peculiar orientation, blazars are very rare. Assuming that the maximum angle 
w.r.t. the line of sight an AGN jet can have for a source to be called a blazar is $\sim$15$^{\circ}$, 
only $\sim$3$\%$ of all radio-loud AGN, and therefore $\sim$0.3$\%$ of all AGN, are blazars. For a 
$\sim$1$\%$ fraction of galaxies hosting an AGN, this implies that only 1 out of $\sim$30,000 
galaxies is a blazar. Hence, the fact that the GeV and TeV skies are dominated by blazars is 
surprising. The explanation stems of course from the blazars' defining characteristics: 
$\bullet$ {\it high-energy electrons}: in some blazars the synchrotron emission peaks in the 
X-ray range: this suggests the presence of high-energy electrons that can produce HE/VHE 
radiation via Compton scattering; 
$\bullet$ {\it strong non-thermal (jet) component}: HE/VHE emission is clearly non-thermal 
and related to the jet: the stronger the latter, the stronger the former; and 
$\bullet$ {\it relativistic beaming}: in sources as compact as blazars (as suggested by their 
short variability timescales) all GeV photons, would be absorbed through pair-producing 
$\gamma\gamma$ collisions with target X-ray photons. Beaming ensures the intrinsic radiation 
density to be much smaller than the observed one, so that VHE photons encounter a much lower 
$\gamma\gamma$ opacity and hence manage to leave the source. Furthermore, relativistic beaming 
causes a strong amplification of the (jet's) observed flux ($f_{\rm obs}$$=$$\delta^p$$f_{\rm obs}$), 
so providing a powerful bias toward blazar detection.

\section{AGILE, MAGIC and blazars}

The GRID instrument onboard the AGILE satellite (\cite{t+00}; launched on April 23, 2007) 
is designed to provide data in the 30 MeV-50 GeV with a sensitivity of $\sim$0.05 Crab 
units at $E$$>$1 GeV, i.e. $\sim$2 times better than EGRET's. Also, w.r.t. the latter, 
AGILE has a larger FOV ($\sim$1/5 of the sky) with a flatter instrumental response across 
the FOV, and a smaller source location radius ($\sim$20 arcmin). 

\begin{figure}
\vspace{4.2cm}
\includegraphics{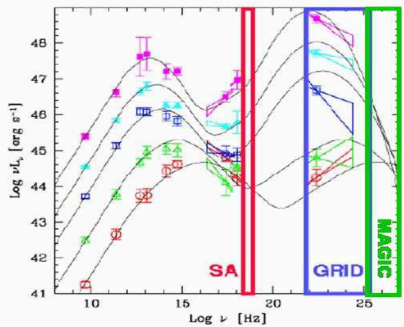}
\caption{The AGILE/GRID, SuperAGILE, and MAGIC bands superimposed on the blazar SED sequence
(adapted from \cite{f+98}).}
\end{figure}

How many blazars AGILE will discover depends on several factors, most notably the blazars' 
spectral properties and duty cycle in the $\gamma$-ray band. To get an estimate, let's 
recall that: 
{\it (i)} $\magcir$130 blazars were detected by EGRET; 
{\it (ii)} AGILE is $\magcir$2 times more sensitive than EGRET; 
{\it (iii)} the EGRET blazar number counts are Euclidean, $N($$>$$S) \propto S^{-1.5}$ (where 
$S$ is the flux): we assume that $S$ is representative of the bolometric flux and the counts 
stay Euclidean down to AGILE's sub-EGRET flux regime. The number of blazars that AGILE may 
detect above 1 GeV is then $\sim$350, of which $\sim$200 new discoveries. However, since at 
$>$100 MeV the sensitivity of AGILE/GRID matches that of EGRET, the actual number of new 
detections could be $\sim$100, i.e. $\sim$20 per FOV (\cite{v+03}).

How many AGILE blazars will be detected by MAGIC, depends on the combined 
instrumental biases convoluted with the intrinsic source properties (see Fig.1). The shape of the 
blazar SED and its trend with luminosity (HBLs/LBLs have lower/higher observed bolometric 
luminosities \cite{f+98,g+98}) causes AGILE and MAGIC to be biased toward detecting, 
respectively, LBLs and HBLs. Indeed, AGILE/GRID will sample the trough between the two 
(synchrotron and Compton) humps in HBLs, and the Compton peak in LBLs -- this makes LBLs 
better targets for AGILE. (In fact, EGRET detected very few HBLs.) MAGIC (\cite{l04}), on the 
other hand, samples the steeply falling side of the Compton hump in LBLs, and the Compton 
peak in HBLs -- this makes HBLs better targets for MAGIC. (In fact, 15/16 of the established 
TeV blazars are HBLs.) 
 
We conclude that the probability of a joint AGILE-MAGIC detection is highest for: 1) blazars 
whose SED properties are intermediate between LBLs (AGILE bias) and HBL (MAGIC bias); and 2) 
high-luminosity LBLs, whose slope stays approximately the same in the HE and VHE bands (the 
AGILE/GRID and MAGIC bands are adjacent, and in both the threshold for a 5$\sigma$ detection 
with a 50 hr observation is $\sim$0.05 Crab units). 

To get a clue on the type (LBL vs HBL) of an AGILE/GRID-detected blazar, simultaneous 
data can be used from AGILE's other instrument, the hard-X-ray monitor SuperAGILE. This 
will allow us to check whether its working band, $\sim$15-45 keV, probes the falling part 
of the synchrotron bump (typical of HBLs) or the rising part of the Compton bump (typical 
of LBLs) in the trough between the two humps. 

As a final remark, we should notice that even established VHE blazars, that however have so far 
escaped HE (i.e., EGRET) detection, could in principle be detected by AGILE. For example, 
several of the candidate TeV blazars proposed by \cite{cg02}, for which there are now many TeV 
detections (by MAGIC and H.E.S.S.), may be detected in the HE band where the EGRET 
detections are so few. If so, a HE detection of a blazar would still warrant a new VHE 
observation to ensure simultaneous broadband $\gamma$-ray coverage of its Compton hump.

\section{Summary} 

Blazars, even though representing a tiny minority of galaxies, dominate the 
$\gamma$-ray sky. AGILE will likely detect several hundred blazars, with a 
bias for the LBL type. Triggered by AGILE/GRID HE detections, MAGIC will perform 
follow-up VHE observations to achieve a combined broadband $\gamma$-ray 
detection. However MAGIC is, like all IACTs, biased to detecting blazars of 
the HBL type. Therefore the probability of a joint AGILE-MAGIC detection will 
be maximized, for a given HE flux, for sources that are spectrally intermediate 
between LBLs and HBLs. Additional information from the hard-X-ray monitor 
SuperAGILE will be instrumental to determine the type (LBL, HBL, or intermediate) 
of the EGRET-detected blazar.

\end{document}